\documentclass[floatfix,letterpaper,showpacs,preprint,preprintnumbers,superscriptaddress,prl]{revtex4-1}

\usepackage{amsmath, amssymb}
\usepackage{graphicx}% Include figure files
\usepackage{times}
\usepackage{dcolumn}

\begin{document}

%\preprint{ver.11}

\title{Landau Renormalizations of Superfluid Density in the Heavy Fermion Superconductor CeCoIn$_{5}$}

\author{Lei Shu}
\affiliation{State Key Laboratory of Surface Physics, Department of Physics, Fudan University, Shanghai 200433, People's Republic of China}
\author{D. E. MacLaughlin}
\author{C. M. Varma}
\affiliation{Department of Physics and Astronomy, University of California, Riverside, California 92521, USA}
\author{O. O. Bernal}
\affiliation{Department of Physics and Astronomy, California State University, Los Angeles, California 90032, USA}
\author{P.-C. Ho}
\author{R. H. Fukuda}
\affiliation{Department of Physics, California State University, Fresno, California 93740,USA}
\author{X. P. Shen}
\affiliation{State Key Laboratory of Surface Physics, Department of Physics, Fudan University, Shanghai 200433, People's Republic of China}
\author{M. B. Maple}
\affiliation{Department of Physics, University of California, San Diego, La Jolla, California 92093,USA}

\date{\today}

\begin{abstract}
The formation of heavy fermion bands can occur by means of the conversion of a periodic array of local moments into itinerant electrons via the Kondo effect and the huge consequent Fermi-liquid renormalizations. Leggett predicted for liquid $^3$He that Fermi-liquid renormalizations change in the superconducting state, leading to a temperature dependence of the London penetration depth~$\Lambda$ quite different from that in the BCS theory. Using Leggett's theory, as modified for heavy fermions, it is possible to extract from the measured temperature dependence of $\Lambda$ in high quality samples both Landau parameters $F_0^s$ and $F_1^s$; this has never been accomplished before. A modification of the temperature dependence of the specific heat $C_\mathrm{el}$, related to that of $\Lambda$, is also expected. We have carefully determined the magnitude and temperature dependence of $\Lambda$ in CeCoIn$_5$ by muon spin relaxation rate measurements to obtain $F_0^s = 36 \pm 1$ and $F_1^s = 1.2 \pm 0.3$, and find a consistent change in the temperature dependence of electronic specific heat $C_\mathrm{el}$. This, the first determination of $F_1^s$ with a value~$\ll F_0^s$ in a heavy fermion compound, tests the basic assumption of the theory of heavy fermions, that the frequency dependence of the self-energy is much more important than its momentum dependence.
\end{abstract}

\pacs{71.27.+a, 74.70.Tx, 75.40.-s,76.75.+i}% PACS, the Physics and Astronomy
 % Classification Scheme.

\maketitle

A major development in condensed-matter physics over the last 35 years has been the discovery and investigation of heavy-fermion compounds and the unconventional superconductivity they exhibit~\cite{Stewart84,Grewe91}. For this class of systems, based on rare-earth and actinide elements (i.e., elements with partially filled 4$f$ or 5$f$ electron shells), the attribute ``heavy" is often associated with the Kondo effect (reflecting the correlation between localized $f$ moments and conduction electrons) which leads to strong renormalization of the effective mass of the electron at low temperatures. Heavy-fermion compounds behave like a system of heavy itinerant electrons, the properties of which can be described in the framework of a Landau Fermi-liquid formalism.

Leggett~\cite{Leggett65} predicted for liquid $^3$He that Fermi-liquid renormalizations change in the superconducting state, leading to temperature dependences of physically observable quantities, e.g., the London penetration depth $\Lambda$, quite different from those in the BCS theory. For $T \rightarrow T_c$, $\Lambda$ is renormalized from its value in the BCS theory by the effective mass, but for $T \rightarrow 0$, where there are no thermally excited quasiparticles, $\Lambda$ retains the BCS value. A modification of the temperature dependence of the electronic specific heat $C_\mathrm{el}$ related to that of $\Lambda$ is also expected.

Unlike liquid $^3$He, heavy fermions are two-component systems in which Landau renormalization, which relies on Galilean invariance, does not work. In heavy-fermion systems the effective mass is primarily determined by the ``compressibility renormalization coefficient" $F_0^s$, with $F_1^s$ only a correction to it~\cite{Varma85}. Then Leggett's change of renormalization on entering the superconducting state, which only depends on $F_1^s$, is modified. Using Leggett's theory, as modified for heavy-fermions~\cite{Varma86}, it is possible to extract both the Landau parameters $F_0^s$ and $F_1^s$ from the measured temperature dependence of the London penetration depth in high quality samples. This has, however, never been accomplished before.

This Letter reports muon spin relaxation ($\mu$SR) experiments in the superconducting and normal states of CeCoIn$_5$, and discusses their implications for the theory of heavy-fermion systems. The London penetration depth derived from the magnetic field distribution in the vortex lattice is shown to exhibit an unusual temperature dependence that is nevertheless consistent with the temperature dependence of $C_\mathrm{el}$; this is a strong check on the experimental results. The Landau parameters $F_0^s$ and $F_1^s$ are obtained for CeCoIn$_5$, and obey the key strong inequality~$F_0^s \gg F_1^s$.

Single crystals of CeCoIn$_5$ were synthesized by means of an indium self-flux method~\cite{Fisk89}, centrifuged, and etched in HCl solution to remove the excess indium. Thin plate-like single crystals were obtained with large faces corresponding to the (001) basal plane. The crystals were aligned and glued to a silver holder covering $10\times10$ mm$^2$ using dilute GE varnish. $\mu$SR experiments were performed on the M15 beam line of TRIUMF, Vancouver, Canada. A top-loading-type dilution refrigerator was used to cool the specimen down to 16 mK.

Transverse-field $\mu$SR (TF-$\mu$SR) has been used extensively to study the vortex state of type-II superconductors~\cite{Sonier00,Sonier07}. In a TF-$\mu$SR experiment, spin-polarized positive muons with a momentum of 29 MeV/c are implanted one at a time into a sample in an external magnetic field $\mu_0 H$ (field cooled from above $T_c$ in a superconductor) applied perpendicular to the initial muon spin polarization. Each muon precesses around the local field $B$ at its site at the Larmor frequency $\omega = \gamma_\mu B$, where $\gamma_\mu / 2\pi = 135.5342$ MHz/T is the muon gyromagnetic ratio. On decay of the muon after an average lifetime of 2.2 $\mu$s, a positron is emitted preferentially along the direction of the muon spin. The time evolution of the muon spin polarization is determined by detecting decay positrons from an ensemble of 1--$2\times 10^7$ muons.

The functional form of the muon spin polarization depends on the field distribution. In the mixed state of a type-II superconductor, the applied magnetic field induces a flux-line lattice (FLL), where the internal magnetic field distribution is determined by the magnetic penetration depth, the vortex core radius, and the structure of the FLL\@. The muon spin relaxation rate is related to the rms width $[\overline{(\Delta B)^2}]^{1/2}$ of the internal magnetic field distribution in the FLL\@. In turn, $[\overline{(\Delta B)^2}]^{1/2}$ is proportional to $\Lambda^{-2}(T)$, the properties of which we discuss after presenting the experimental results.

\begin{figure}[ht]
 \begin{center}
 \includegraphics[width=0.45\textwidth]{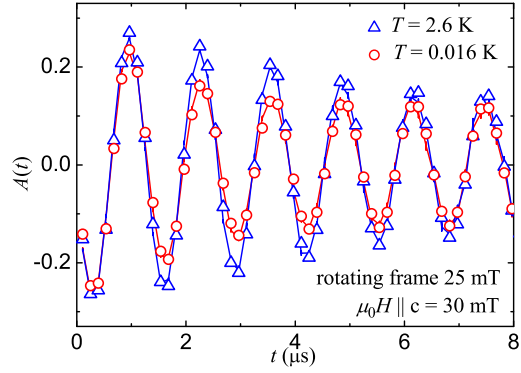}
 \caption{(Color online) TF-$\mu$SR asymmetry spectra $A(t)$ of CeCoIn$_{5}$ in the normal (triangles) and superconducting (circles) states, taken at $\mu_0H= 30$ mT for $H\,\|\, c$. Curves represent fits to the data (see text). The data are shown in a frame rotating at a frequency corresponding to a field of 25~mT~\cite{Brewer94}.}
 \label{fig:Asy}
 \end{center}
\end{figure}

Representative TF-$\mu$SR muon-spin precession signals at an applied field of 30 mT are shown in Fig.~\ref{fig:Asy}
in the normal and superconducting states of CeCoIn$_5$. The $\mu$SR asymmetry spectrum consists of two contributions: a signal from muons that stop in the sample, and a slowly-relaxing background signal from muons that stop in the silver sample holder. %These contributions are clearly seen in Fig.~\ref{fig:Asy}. At short times the sample contribution dominates.
As seen in Fig.~\ref{fig:Asy}, in the superconducting state the damping of the signal is enhanced at early times due to the field broadening generated by the FLL\@. For times longer than ${\sim} 6~\mu$s in the normal state and ${\sim} 4~\mu$s in the superconducting state, only the background signal persists.

The $\mu$SR asymmetry spectra in CeCoIn$_5$ are well described by the fitting function
\begin{equation}
\begin{split}
 \label{eq:Asy}
 A(t) = & A_0\left[f_s\exp(-{\textstyle\frac{1}{2}}\sigma_s^2t^2)\cos(\omega_s t+\phi)\right.\\
 & + \left.(1-f_s)\exp(-{\textstyle\frac{1}{2}}\sigma_b^2t^2)\cos(\omega_b t+\phi)\right],
 \end{split}
 \end{equation}
where $A_0$ is the initial asymmetry of the signal and $f_s$ denotes the fraction of muons stopping in the sample. The Gaussian relaxation rate~$\sigma_s$ from the sample is due to nuclear dipolar fields in the normal state, and is enhanced in the superconducting state by the FLL field inhomogeneity. The precession frequency~$\omega_s$ is reduced due to diamagnetic screening. The background relaxation rate~$\sigma_b$ is negligibly small, and the initial phase~$\phi$ and background frequency~$\omega_b$ are constant. The curves in Fig.~\ref{fig:Asy} are fits of Eq.~(\ref{eq:Asy}) to the data.

\begin{figure}[ht]
 \begin{center}
 \includegraphics[clip=,width=0.45\textwidth]{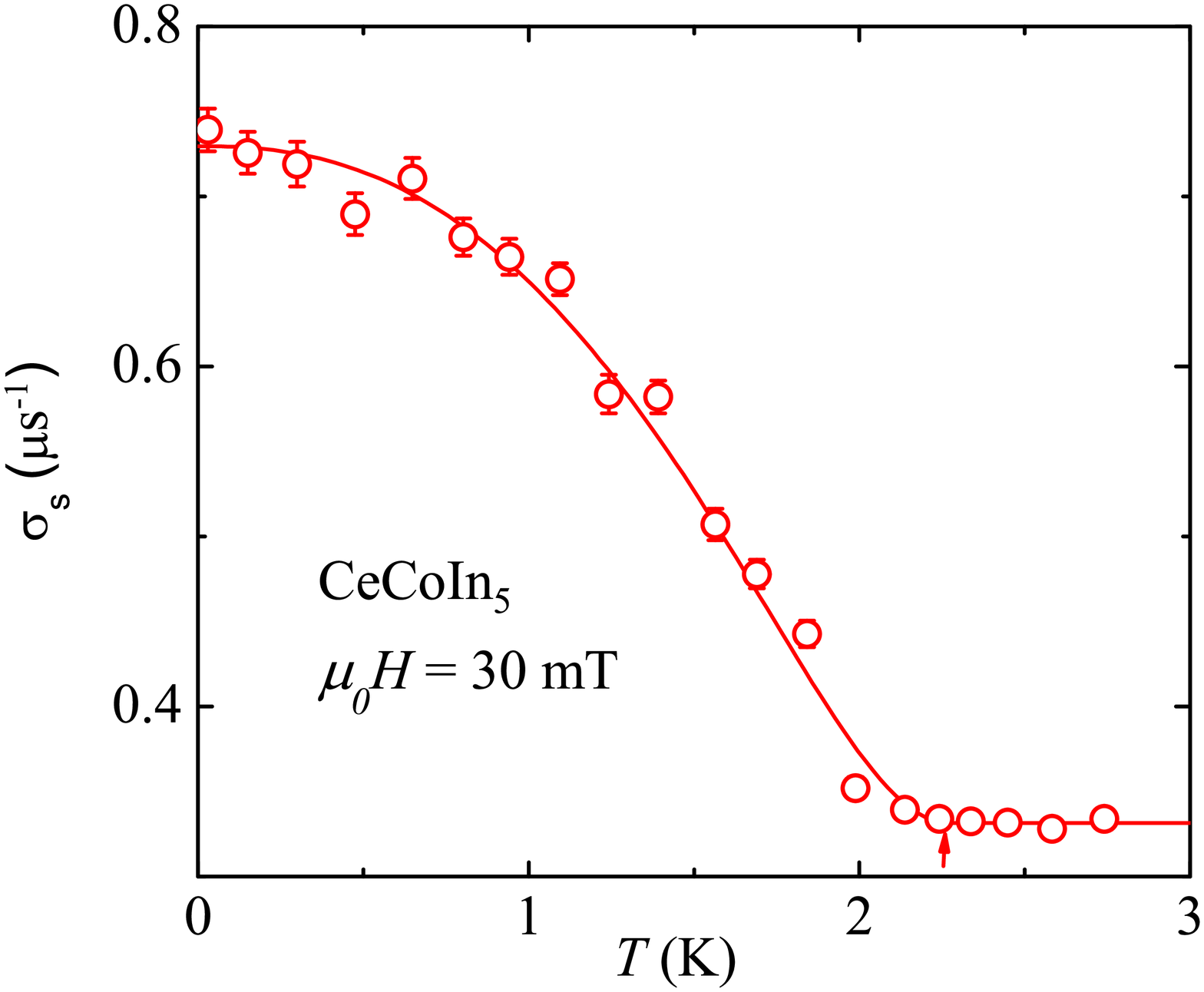}
 \caption{(Color online) Temperature dependence of the muon Gaussian relaxation rate~$\sigma_s$ in CeCoIn$_{5}$ under a transverse magnetic field $\mu_0H= 30$~mT for $H\| c$. Solid lines% are
: power-law fits to the data as described in the text. Arrow: $T_c$.}
 \label{fig:Rlx}
 \end{center}
\end{figure}

The temperature dependence of $\sigma_s$for CeCoIn$_5$ is shown in Fig.~\ref{fig:Rlx}.
Salient features of these data are the temperature independence of $\sigma_s$ above the transition temperature $T_c$, and the increase of $\sigma_s$ with decreasing temperature below $T_c$. This result indicates that bulk superconductivity occurs below $T_c$, consistent with electrical resistivity and specific heat measurements~\cite{Petrovic01a,Shu11}.

The internal field distribution in the vortex state is the convolution of the field distributions due to the vortex lattice and the nuclear dipolar field distribution of the host material, leading to
 \begin{equation}
 \label{eq:sigma}
 \sigma_s^2=\sigma_\mathrm{FLL}^2+\sigma_{\rm dip}^2.
 \end{equation}
 $\sigma_{\rm dip}^2$ is temperature-independent in the normal state, and is not expected to change in the superconducting state. After determining $\sigma_{\rm dip}^2 = 0.33 \mu s^{-1}$ from the normal-state data, we concentrate on $\sigma_\mathrm{FLL}$ and its relation to $\Lambda$.

The temperature dependence of $\sigma_\mathrm{FLL}$ can be fit with
\begin{equation}
 \label{eq:sigmaFLL}
 \sigma_\mathrm{FLL}(T)=\sigma_\mathrm{FLL}(0)\left[1-(T/T_c)^n\right] \quad (T < T_c),
 \end{equation}
with the fitting parameters $\sigma_\mathrm{FLL}(0) = 0.65(2) \mu s^{-1}$, $T_c=2.27(2)K$, and $n=2.4(2)$. The value of $T_c$ is consistent with transport measurements~\cite{Petrovic01a,Shu11}. In an isotropic extreme type-II superconductor, the second moment $ \overline{(\Delta B)^2}$ is approximately given by $ \overline{(\Delta B)^2}=\sigma_\mathrm{FLL}^2/\gamma_{\mu}^2=0.00371\Phi_0^2\Lambda^{-4}$~\cite{Brandt88}, where $\Phi_0$ is the flux quantum. The penetration depth $\Lambda(0)$ is then obtained from $\sigma_\mathrm{FLL}(0)$ to be 406(12) nm \footnote{Howald \textit{et al.}, Phys. Rev. Lett. {\bf 110}, 017005 (2013), report $\Lambda_a(0)= 350(12)$~nm, which is close to the present value. Higemoto \textit{et al.}, J. Phys. Soc. Japan {\bf 71}, 1023 (2002), using a larger magnetic field $\mu_0H=0.3$~T, report $\Lambda_{ab}(0) = 550$~nm for CeCoIn$_5$ with a slightly lower $T_c= 2.0 $ K\@. In this case the fitting process might have been difficult due to vortex-core effects in the denser FLL~\protect\cite{Sonier00, Sonier07}. The discrepancy may also arise from magnetic-field effects and pair-breaking disorder in the lower $T_c$ samples. As noted by Howald \textit{et al.}, a number of effects (multiple muon stopping sites, vortex core corrections) can affect measurements of the penetration depth at high fields.
}.

The London penetration depth is related to the superfluid density $\rho_s(T)$ by
\begin{equation}
\label{Lambda1}
\Lambda^{-2} = \frac{4\pi e^2 \rho_s(T)}{m_dc^2},
\end{equation}
where $m_d$ is the dynamical mass, defined as the ratio between the carriers' momentum and velocity, in the limit $T \to 0$. \textit{In the pure limit and for a single band}, $\Lambda^{-2}$ may be written in terms of its value at $T \to 0$, which is the pure diamagnetic contribution, and the paramagnetic temperature-dependent contribution $K(T)$ due to the depletion of the condensate by the thermal excitation of quasiparticles. Leggett pointed out that the latter is renormalized by the Landau parameter $F_1^s$ such that
\begin{equation}
\label{Lambda2}
\Lambda^{-2} = \frac{4\pi e^2}{c^2}\frac{N}{m_d}\left[1 - K(T)\right],
\end{equation}
where $N$ is the carrier density,
\begin{equation}
\label{KT}
K(T) = \frac{\left(1+ \frac{1}{3} F_1^s\right)Y(T)}{1 + \frac{1}{3} F_1^sY(T)},
\end{equation}
and $Y(T)$ is the Yosida function
\begin{equation}
Y(T) = -N(\epsilon_F)^{-1}\sum_\mathbf{k} df/dE_\mathbf{k};
\end{equation}
here $N(\epsilon_F)$ is the density of states at the Fermi energy and $f(E_\mathbf{k})$ is the usual Fermi distribution function. Leggett's theory is formulated for a Galilean invariant system such as liquid $^3$He in which the mass $m_d$ must remain unrenormalized: $m_d = m$, the bare mass. It was pointed out that for heavy-fermion compounds, which are mutually interacting multi-component systems, the heavy electrons come from the renormalization of $f$ moments to itinerant electrons by the Kondo effect~\cite{Varma85}, through exchange interactions with the $s$, $p$ and $d$ bands. Then $m_d$ is given by the renormalization of the quasiparticle amplitude such that~\cite{Varma86}
\begin{equation}
\label{eq:meff}
m_d \approx m(1+F_0^s).
\end{equation}
The basic assumption behind this relationship and the strong Landau-parameter inequality~$F_0^s \gg F_\ell^{s,a}$ for $\ell \geqslant 1$ is that for heavy fermions the frequency dependence of the self-energy is much more important than the momentum dependence~\cite{Varma85,Varma86}. This assumption also underlies the theory of heavy fermions through various theoretical advances, such as slave boson methods~\cite{Gabriel86} and dynamical mean-field theory~\cite{Georges96}. We will see below how these assumptions are tested by the experimental results presented here.

In a multi-band situation, one should in principle have instead of the factor $N/m_d$ the sum of contributions from all the bands, but since $F_0^s \gg 1$ for heavy fermions, the heavy band contribution dominates in the determination of $\Lambda$ (and $C_\mathrm{el}$), and to a good approximation only the heavy bands need to be considered.

For $d$-wave superconductivity, for which there is overwhelming evidence in CeCoIn$_5$~\cite{Petrovic01a,Izawa01,Settai01,Movshovich01,Aoki04,Tanatar05,Stock08,An10}, $Y(T) \propto (T/T_c)^3$ for three-dimensional materials in the pure limit, for a state with line nodes of the gap function and magnetic field perpendicular to the line nodes. Without change of renormalizations in the superconducting state, $\Lambda^{-2}(T) \propto 1 -(T/T_c)^3$ would therefore be expected. This is not observed in the experiments% (X XX Fig.~\ref{fig:Lam})
; the best fit to the data yields an exponent of $2.4 \pm 0.2$ ).

Before proceeding further, we ascertain that CeCoIn$_5$ is indeed in the pure limit~\cite{Movshovich01,Petrovic02,Bianchi03,Martin05}; i.e., the mean-free path is much larger than the superconducting coherence length, which is determined by the superconducting gap and the renormalized Fermi velocity to be about 5~nm~\cite{Martin05}. For the sample of CeCoIn$_5$ studied in our experiments, the extrapolated normal state resistivity is only a few $\mu \Omega \cdot$cm~\cite{Shu11}, which gives a mean free path of more than 50~nm.

\begin{figure}[ht]
 \begin{center}
 \includegraphics[clip=,width=0.45\textwidth]{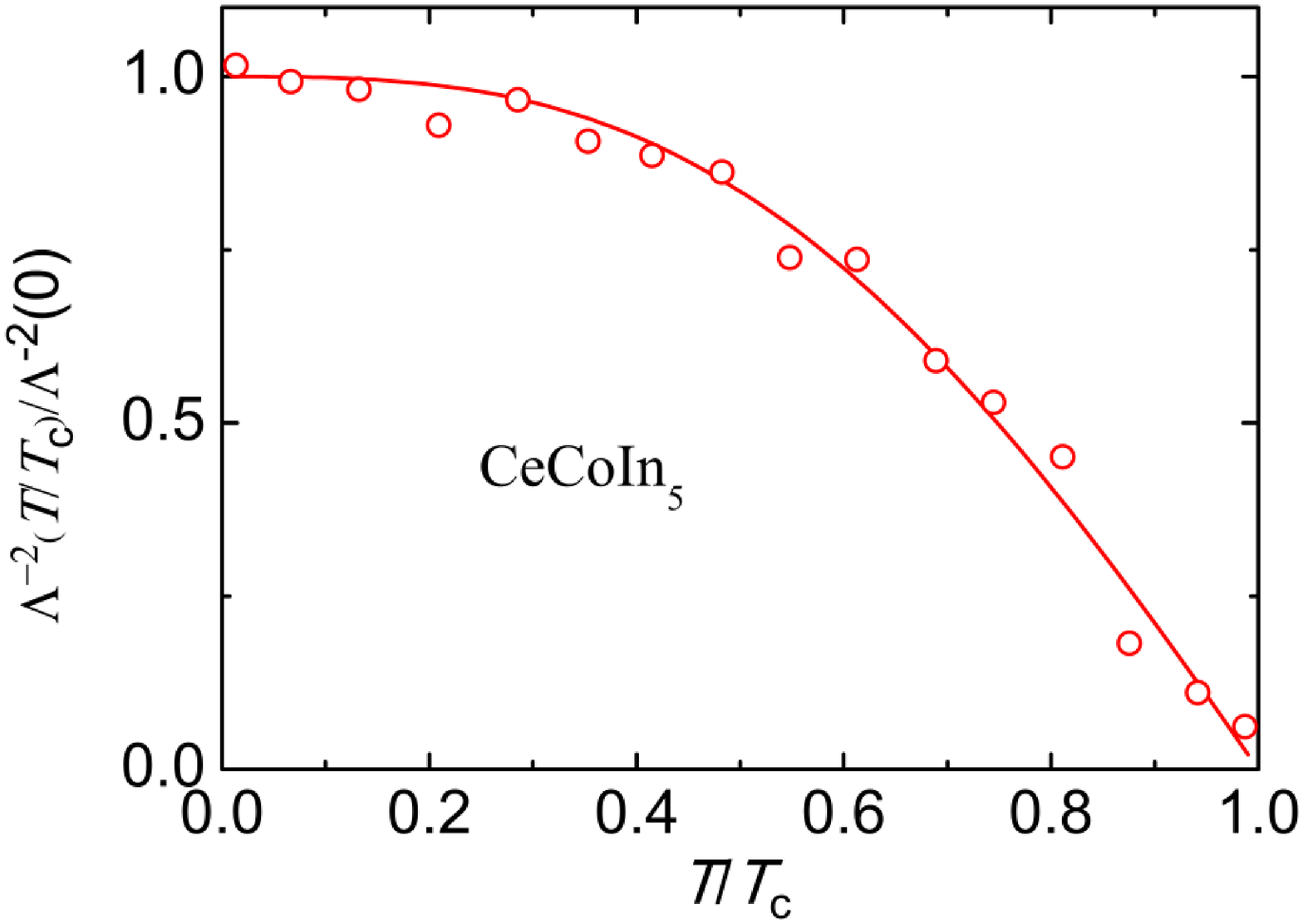}
 \caption{(Color online) Superfluid density $\Lambda^{-2}(T/T_c)/\Lambda^{-2}(0)$ as a function of normalized temperature $T/T_c$ in CeCoIn$_{5}$. Curve: fit of Eq.~(\ref{eq:Lambda3}) to the data. }
 \label{fig:Lam}
 \end{center}
\end{figure}

Using Eqs.~(\ref{Lambda2}) and (\ref{KT}), we can write
\begin{equation} \label{eq:Lambda3}
\frac{\Lambda^{-2}(T/T_c)}{\Lambda^{-2}(0)}=1-\frac{\left(1+\frac{1}{3}F_1^s\right)(T/T_c)^3}{1+\frac{1}{3}F_1^s(T/T_c)^3}.
\end{equation}
Figure~\ref{fig:Lam}
presents the fit of Eq.~(\ref{eq:Lambda3}) to the measured temperature dependence of $ \Lambda^{-2}(T/T_c)/\Lambda^{-2}(0) = \sigma_\mathrm{FLL}(T/T_c)/\sigma_\mathrm{FLL}(0)$ in CeCoIn$_5$, from which we extract $F_1^s = 1.2 \pm 0.3$.

We turn next to the determination of $F_0^s$, which requires knowing the value of $N$. The valence of Ce is known from independent measurements to be $+3$~\cite{Booth11,Dudy13}, thereby contributing one $f$ electron to the heavy conduction band. For CeCoIn$_5$, one then has $N=1/V_\mathrm{cell}$. Using this value, the measured $\Lambda(0) = 406(12)$~nm, and Eq.~(\ref{eq:meff}), we find $F_0^s = 36(1)$.

Independent confirmation of these results is obtained by comparing the predictions for the specific heat with measurements of the electronic specific heat~$C_\mathrm{el}(T)$~\cite{Shu11}. It is hard to calculate the absolute value of $C_\mathrm{el}$. From de Haas-van Alphen measurements~\cite{Settai01,McCollam05} we know that there are at least three sheets of the Fermi surface with varying masses and areas. To get absolute values, in addition to the parameters $F_0^s$ and $F_1^s$, we need the details of the dispersion relations of all the bands that cross the Fermi surface and the relative contribution of the $f$-electrons to the bands since they alone are affected by the strong renormalizations. If we make the simplest assumption of a parabolic heavy band, from the relation $C_\mathrm{el} /T= \pi^2 N k_B^2 m_\mathrm{eff}/\hbar^2k_F^2$, where the Fermi wave vector is given by $k_F=(3\pi^2 N)^{1/3}$ and $m_\mathrm{eff} = m_d\left(1 + \frac{1}{3}F_1^s \right)$~\cite{Varma86}, we obtain $C_\mathrm{el}/T \sim 148$~mJ/mol~K$^2$, compared with the experimental value of $\sim$300~mJ/mol~K$^2$ at $T=T_c$~\cite{Petrovic01a,Shu11}. Considering the simplified band assumption, this may be regarded as successful.

Finally, we consider the change of $C_\mathrm{el}(T)$ due to the Leggett renormalizations. An approximate $T^3$ dependence is observed~\cite{Petrovic01a,Movshovich01,Ikeda01,Tanatar05,An10,Shu11} for $0.2 \lesssim T/T_c < 1$~\footnote{After subtraction of a nuclear Schottky contribution, the measured specific heat varies as $\gamma_0T + aT^2$ below $T/T_c \sim 0.2$~\protect\cite{Petrovic01a,Movshovich01,Ikeda01,Tanatar05,An10}. We do not consider the ${\sim}T^3$ renormalization correction in this region, since it is very small [$\lesssim 1$\% from Eq.~({\protect\ref{eq:spht}})].}. Nonzero $F_1^s$ renormalizes $C_\mathrm{el}(T)$ in the superconducting state by an amount proportional to the normal fluid density $\rho_n(T) = 1 - \Lambda^{-2}(T/T_c)/\Lambda^{-2}(0)$. In the pure limit this leads to
\begin{equation}
\label{eq:spht}
C_\mathrm{el}(T) \propto \left[1+{\textstyle\frac{1}{3}}F_1^s \frac{(1+\frac{1}{3}F_1^s)(T/T_c)^3}{1+\frac{1}{3}F_1^s(T/T_c)^3} \right] \left(\frac{T}{T_c}\right)^3
\end{equation}

\begin{figure}[ht]
 \includegraphics[clip=,width=0.45\textwidth]{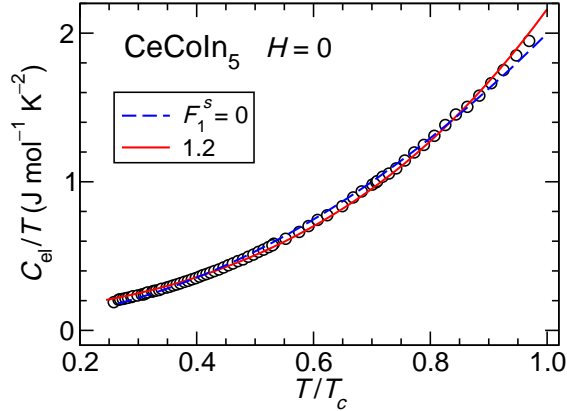}
 \caption{(Color online) Dependence of specific heat divided by temperature $C_\mathrm{el}(T)/T$ on normalized temperature $T/T_c$ in CeCoIn$_{5}$. Data from Ref.~\protect\cite{Shu11}. Curves: fits of Eq.~(\ref{eq:spht}) (plus small offsets~\cite{Note2}) for $F_1^s =  0$ and 1.2 to the data. The fits and data are nearly indistinguishable except near $T_c$}
 \label{fig:Cel}
\end{figure}

This is tested by fits of $C_\mathrm{el}(T)/T$ from Eq.~(\ref{eq:spht}) (plus small offsets~\cite{Note2}) to the data~\cite{Shu11} for $F_{1}^s = 0$ and 1.2, shown in Fig.~\ref{fig:Cel}.
The consistency of the results is confirmed: the specific heat, unlike the penetration depth, is quite insensitive to the value of $F_{1}^s$. This can also be seen by expanding Eqs.~(\ref{eq:Lambda3}) and (\ref{eq:spht}) for $\Lambda^{-2}(T)/\Lambda^{-2}(0)-1$ and $C_\mathrm{el}(T)/C_\mathrm{el}(T_c)$ in $F_1^s$; the leading terms in the corrections are of first and second order, respectively.

In conclusion, we have determined the Fermi-liquid parameters~$F_0^s$ and $F_1^s$ from $\mu$SR measurements of the penetration depth in CeCoIn$_5$. This is the first such determination of both parameters. The inequality~$F_1^s \ll F_0^s$ is fulfilled, thereby verifying a basic assumption in the theory of heavy fermions.

\begin{acknowledgments}
 This research is supported by the National Natural Science Foundation of China (11204041), Natural Science Foundation of Shanghai, China (12ZR1401200), Research Fund for the Doctoral Program of Higher Education of China (2012007112003), the overseas Returnees Start-Up Research Fund of the Ministry of Education in China, and by the U.S. Department of Energy (DOE) under Research Grant No.~DE-FG02-04ER46105 (sample synthesis at UC San Diego) and by the U.S. National Science Foundation under Grant Nos.~DMR 0802478 (sample characterization at UC San Diego), DMR 0801407 and 1206298 (UC Riverside), DMR 1104544 (CSU Fresno), and DMR 1105380 (CSU Los Angeles).
\end{acknowledgments}

%\bibliography{./115}

%Merlin.mbs v4.21 2009-07-09.
%

\end{document}